\documentclass[12pt]{article}
\usepackage{latexsym}
\usepackage{amsmath,,calrsfs}
\usepackage{amsfonts}
\usepackage{amssymb}
\usepackage{amscd}
\usepackage{bbm}
\usepackage{fancybox}
\usepackage{cite}
\usepackage{amsmath,amsfonts,amsbsy}
\usepackage{pstricks,pst-node}
\usepackage[small,bf,hang]{caption2}
\usepackage{graphicx}
\usepackage{epsfig}
\usepackage{psfrag}
\usepackage{comment}
\usepackage{youngtab}
\usepackage{young}

\usepackage{mathrsfs}
\usepackage{array}
\usepackage{graphicx}
\usepackage{slashed}

\newcommand{\Z}{\mathscr{Z}}

\usepackage{float}

\psset{unit=1.3cm,linewidth=.5pt,radius=.2}  

\usepackage{multirow}                     
\usepackage{float}                          
\usepackage{lscape}                         
\usepackage{bm}

\newcommand{\rmd}{\mathrm{d}}

\newcommand{\beq}{\begin{equation}}
\newcommand{\eeq}{\end{equation}}
\newcommand{\bi}{\begin{itemize}}
\newcommand{\ei}{\end{itemize}}
\newcommand{\bt}{\begin{tabular}}
\newcommand{\et}{\end{tabular}}
\newcommand{\bc}{\begin{center}}
\newcommand{\ec}{\end{center}}

\newcommand{\be}{\begin{equation}}
\newcommand{\ee}{\end{equation}}
\newcommand{\bea}{\begin{eqnarray}}
\newcommand{\eea}{\end{eqnarray}}
\newcommand{\ba}{\begin{array}}
\newcommand{\ea}{\end{array}}

\def\bbox{{\,\lower0.9pt\vbox{\hrule \hbox{\vrule height 0.2 cm
\hskip 0.2 cm \vrule height 0.2 cm}\hrule}\,}}
\newcommand{\dsl}{\pa \kern-0.5em /}

\makeatletter \@addtoreset{equation}{section} \makeatother

\def\slashchar#1{\setbox0=\hbox{$#1$}           
   \dimen0=\wd0                                 
   \setbox1=\hbox{/} \dimen1=\wd1               
   \ifdim\dimen0>\dimen1                        
      \rlap{\hbox to \dimen0{\hfil/\hfil}}      
      #1                                        
   \else                                        
      \rlap{\hbox to \dimen1{\hfil$#1$\hfil}}   
      /                                         
   \fi}

\newcommand{\rmi}{\mathrm{i}}

\begin{document}

\begin{titlepage}
\begin{center}

\rightline{ UG-13-66}
\rightline{ MIFPA-13-xx}
\rightline{ TUW-13-07}

\vskip 1.5cm

{\Large \bf  Newton-Cartan Supergravity }

\vskip 1cm

{\bf Roel Andringa\,${}^1$\,, Eric A.~Bergshoeff\,${}^1$\,,
Jan Rosseel\,${}^2$ and  Ergin Sezgin\,${}^3$}\\

\vskip 25pt

{\em $^1$ \hskip -.1truecm
\em Centre for Theoretical Physics,
University of Groningen, \\ Nijenborgh 4, 9747 AG Groningen, The
Netherlands \vskip 5pt }

{email: {\tt R.Andringa@rug.nl; E.A.Bergshoeff@rug.nl
}} \\

\vskip .4truecm

{\em $^2$ \hskip -.1truecm
\em Institute for Theoretical Physics,
Vienna University of Technology \\
Wiedner Hauptstr. 8-10/136 \\
A-1040 Vienna, Austria
\vskip 5pt }

{email: {\tt rosseelj@hep.itp.tuwien.ac.at}} \\

\vskip .4truecm

{\em $^3$ \hskip -.1truecm
\em George P. \& Cynthia Woods Mitchell Institute for \\ Fundamental Physics and Astronomy,
Texas A\&M University, \\ College Station, TX 77843, USA
\vskip 5pt }

{email: {\tt sezgin@physics.tamu.edu}} \\

\end{center}

\vskip 0.5cm

\begin{center} {\bf ABSTRACT}\\[3ex]
\end{center}

We construct a supersymmetric extension of three-dimensional
Newton-Cartan gravity by gauging a super-Bargmann algebra. In order
to obtain a non-trivial supersymmetric extension of the Bargmann
algebra one needs at least two supersymmetries leading to a
$\mathcal{N}=2$ super-Bargmann algebra. Due to the fact that there
is a universal Newtonian time, only one of the two supersymmetries
can be gauged.
The other supersymmetry
is realized as  a fermionic Stueckelberg symmetry and only survives as a global supersymmetry. We explicitly show how,
in the frame of a Galilean observer, the system reduces to a
supersymmetric extension of the Newton potential. The corresponding supersymmetry rules can only be defined, provided we also introduce a `dual Newton potential'.
We comment on the four-dimensional case.

\end{titlepage}

\newpage
\setcounter{page}{1} \tableofcontents

\newpage

\section{Introduction}

It is known that non-relativistic Newtonian gravity can be reformulated in a geometric way, invariant under general coordinate transformations, thus mimicking General Relativity. This reformulation is known as Newton-Cartan theory \cite{Cartan,MTW}. By (partially) gauge fixing general coordinate transformations, non-geometric formulations can be obtained. The extreme case is the one in which one gauge fixes such that one only retains the Galilei symmetries, corresponding to a description in free-falling frames, in which there is no gravitational force. A less extreme case is obtained by gauge fixing such that one not only considers free-falling frames, but also includes frames that are accelerated, with an arbitrary time-dependent acceleration, with respect to a free-falling frame. The observers in such a frame are called `Galilean observers' \cite{Kuchar:1980tw,DePietri:1994je} and the corresponding formulation of non-relativistic gravity is called `Galilean gravity' \footnote{The case in which constant accelerations are considered, instead of time-dependent ones, leads to ordinary Newtonian gravity, described by a time-independent Newton potential.}. In such a frame, the gravitational force is described by the Newton potential $\Phi$. Such frames are related to each other by the so-called `acceleration extended' Galilei symmetries, consisting of an extension of the Galilei symmetries in which constant spatial translations become time-dependent ones\,\footnote{The group of acceleration-extended Galilei symmetries is also called the Milne group
\cite{Duval:1993pe}.}. In this paper, we will construct a supersymmetric version of both Newton-Cartan gravity, as well as Galilean gravity, and show how they are related via a partial gauge fixing.

In a previous work we showed how  four-dimensional Newton-Cartan
gravity can be obtained by gauging the Bargmann algebra\footnote{The Bargmann algebra does not contain any conformal symmetries. Non-relativistic conformal (super)algebras, and their relation to Newton-Cartan space-time, were investigated in \cite{Duval:2009vt,Duval:1993hs}.} which is a
central extension of the Galilei algebra \cite{Andringa:2010it}. An
important step in this gauging procedure is the imposition of a set
of constraints on the curvatures corresponding to the
algebra \cite{Chamseddine:1976bf}. The purpose of these constraints is to convert the
abstract time and space translations of the Bargmann algebra into
general coordinate transformations. In the relativistic case,
i.e.~when gauging the Poincar\'e algebra, one imposes that the torsion, i.e.~the curvature corresponding to the spacetime translations, vanishes:
\begin{equation}
R_{\mu\nu}{}^a(P)=0\,,\hskip 1.5truecm \mu, a=0,1,2,3\,.
\end{equation}
These constraints are called conventional constraints. The same set of
constraints serves another purpose: it can be used to solve for the
spin-connection fields corresponding to the Lorentz transformations
in terms of the other gauge fields. This is different from the
non-relativistic case where  setting the curvature corresponding to time translations equal to zero
is a true constraint:
\begin{equation}\label{bc}
R_{\mu\nu}(H)=\partial_\mu\tau_\nu - \partial_\nu\tau_\mu=0\,.
\end{equation}
This constraint cannot be used to solve for any spin connection. Instead,
it allows us to
write the temporal Vierbein $\tau_\mu$ as
\begin{equation}
\tau_\mu(x^\nu) = \partial_\mu \tau(x^\nu)
\end{equation}
for an arbitrary scalar function $\tau(x^\nu)$. One can use the time
reparametrizations to choose this function equal to the absolute
time which foliates the Newtonian space-time:\,\footnote{We use a notation where $\emptyset$ indicates a curved $\mu=0$ index.}
\begin{equation}\label{fix}
\tau(x^\nu) = x^{\slashed 0}\equiv t\,,\hskip 2truecm  \tau_\mu(x^\nu) = \delta_\mu{}^{\slashed 0}\,.
\end{equation}
This can be viewed as a gauge condition that fixes the time reparametrizations with local parameters $\xi^\emptyset(x^\mu)$
to constant time translations:\,\footnote{With the exception of sections 2.1 and 4, we will assume that any parameter, without any spacetime dependence indicated, is constant. This should be contrasted to fields where we do not always indicate the explicit spacetime dependence.}
\begin{equation}
\xi^\emptyset(x^\nu) = \xi^\emptyset\,.
\end{equation}
One also imposes the conventional constraint that the curvature of the spatial translations equals zero:
\begin{equation}\label{first}
R_{\mu\nu}{}^a(P)=0\,,\hskip 1.5truecm \mu=0,1,2,3; \ a=1,2,3\,.
\end{equation}
However, this constraint by itself is not sufficient to solve for both the spin connection fields corresponding to the
spatial translations as well as the spin connection fields corresponding to the  boost transformations. In order to achieve that one needs to extend the Galilei algebra to the Bargmann algebra and impose that the curvature corresponding to the central extension vanishes as well \cite{Andringa:2010it}. Together with \eqref{first} this conventional constraint can be used to solve for all spin-connection fields.
The invariance
of the non-relativistic theory under central charge transformations corresponds to particle number conservation which is indeed a non-relativistic property.

It is the purpose of this work to extend the construction of
\cite{Andringa:2010it} to the supersymmetric case by gauging a
supersymmetric extension of the Bargmann algebra. A $\mathcal{N}=1$ supersymmetric extension of the Bargmann algebra was considered in \cite{Clark:1983ne}. According to this algebra, the anti-commutator of two supercharges leads to a central charge transformation. We
are however primarily interested in a non-trivial supersymmetric extension
in which the anti-commutator of the fermionic generators contains
the generators corresponding to time and space translations. It
turns out that this can only be achieved provided we consider a
$\mathcal{N}=2$ supersymmetric extension of the Bargmann algebra
\cite{Gomis:2004pw}. The analysis of \cite{Gomis:2004pw} also leads
to a realization of this algebra, as global symmetries, on the
embedding coordinates of a non-relativistic superparticle
propagating in a flat Newtonian space-time.

For technical reasons
explained below, we consider from now on only the three-dimensional case. Three-dimensional gravity is interesting by itself, both relativistically as well as non-relativistically. Although the relativistic theory does not have any local degrees of freedom and there is no interaction between static sources, moving particles can still exhibit non-trivial scattering \cite{Deser:1983tn}. In contrast, in the non-relativistic Newtonian theory, there is an attractive gravitational Newton force that goes as the inverse of the distance between point masses. This theory can thus not be viewed as a non-relativistic limit of General Relativity. Indeed, in the latter, there is no attractive force between static sources, while Newton gravity does exhibit such a gravitational attraction. Coming back to the supersymmetric extensions of non-relativistic gravity, we note that supersymmetric extensions of the three-dimensional Bargmann algebra  were considered in
\cite{Lukierski:2006tr}.

When gauging the $\mathcal{N}=2$ super-Bargmann algebra, one must at
some point impose that the super-covariant extension of the bosonic
curvature $R_{\mu\nu}(H)$ equals zero:
\begin{equation}
{\hat R}_{\mu\nu}(H)=0\,.
\end{equation}
This is the supersymmetric generalization of the constraint \eqref{bc}.
We find that under supersymmetry this constraint leads to another
constraint that sets the super-covariant curvature corresponding to
one of the two gravitini, $\psi_{\mu+}$, equal to zero:
\begin{equation}
{\hat \psi}_{\mu\nu+}=0\,.
\end{equation}
In the same way that the
time reparametrizations, up to constant time translations, can be
used to fix the temporal Dreibein according to \eqref{fix}, one may
now use one of the two local supersymmetries, with arbitrary fermionic parameters $\epsilon_+(x^\mu)$, to set the
 $\psi_{\mu+}$ gravitini equal to zero:
 \begin{equation}
 \psi_{\mu+}=0\,.
 \end{equation}
 This gauge choice fixes the local $\epsilon_+$-supersymmetry to constant ones:
 \begin{equation}
 \epsilon_+(x^\mu) = \epsilon_+\,.
 \end{equation}
 The remaining
supersymmetry, with parameters $\epsilon_-(x^\nu)$ can be  non-trivially gauged.
Only the commutator of a constant and a gauged
supersymmetry leads to a (local) spatial translation. We find that the commutator of
two constant supersymmetries leads to a (constant) time translation while the commutator of two gauged supersymmetries leads to a (local) central charge transformation. It turns out that one can only truncate away the global but not the local supersymmetry. This explains why we need at least two supersymmetries to obtain a non-trivial
supersymmetry algebra where the commutator of two supersymmetries gives a translation.

The above paragraph refers to a so-called `full gauging', in which all symmetries are gauged. This leads to a geometric description of Newtonian supergravity, that uses a temporal and spatial dreibein and is invariant under arbitrary general coordinate transformations. This theory can appropriately be called `Newton-Cartan supergravity'. The case in which we consider a description that is only invariant under the acceleration extended Galilei symmetries, is obtained by a `medium gauging' and the corresponding supergravity theory can be called `Galilean supergravity'. In this work, we will obtain the medium gauging from the fully gauged Newton-Cartan supergravity by a partial gauge fixing\,\footnote{The full gauging corresponding to Newton-Cartan gravity and the medium gauging leading to Galilean gravity have been discussed in  \cite{Andringa:2012uz}.}. The Galilean supergravity we thus obtain, contains a field, corresponding to the Newton potential, as well as a fermionic superpartner. The Newton potential of Galilean supergravity replaces the temporal and spatial Dreibeins of Newton-Cartan supergravity. We find that, in order to write down the supersymmetry transformation rules, we also have to introduce a `dual Newton potential'. The Newton potential and its dual can be seen as real and imaginary parts of a meromorphic function, whose singularities indicate the positions of added point-like sources.

All the above arguments are equally valid when gauging the
four-dimensional $\mathcal{N}=2$ super-Bargmann algebra. However, in
the four-dimensional case we are dealing with the additional
complication that in the relativistic case the algebra can only be
closed provided we introduce more fields than the gauge fields
associated to each of the generators of the algebra. To be precise,
the $\mathcal{N}=2$ super-Poincar\'e algebra requires besides the
usual gauge fields the introduction of an extra Abelian gauge field.
In the non-relativistic case, one would expect that, similarly,
extra fields are needed to close the algebra. We have performed the
four-dimensional gauging procedure and verified that it is not
enough to introduce a single Abelian vector field in the
non-relativistic case. More fields are needed and that is what makes
the four-dimensional case more complicated. In the conclusions we will comment on this issue.

This work is organized as follows. As a warming-up exercise, we will first review
in section 2 the gauging, leading to Newton-Cartan gravity, and subsequent gauge fixing, leading to Galilean gravity, in the bosonic case.
In section 3 we present the 3D
$\mathcal{N}=2$ super-Bargmann algebra. In section 4 we perform the
gauging of this algebra, following the procedure outlined for the
bosonic case in \cite{Andringa:2010it} and reviewed in section 2. We explicitly perform the
gauge fixing that brings us to the frame of a Galilean observer in section 5 and
show how the Newton-Cartan supergravity theory reduces to a
Galilean supergravity theory in terms of a Newton potential and its supersymmetric partner.
We present our
conclusions in section 6. The notation and conventions we use in
this work are presented in a separate appendix.

\section{Newton-Cartan and Galilean gravity}

In this section, we will show how the Newton-Cartan theory can be obtained by gauging the Bargmann algebra and how subsequently Galilean gravity can be obtained by partial gauge fixing.

\subsection{Newton-Cartan gravity}

Our starting point is the Bargmann algebra which is a central extension of the Galilei algebra:
\begin{alignat}{2}\label{Bargmann}
[J_{ab},P_{c}]  &= -2\delta_{c[a}P_{b]}\,, &\hskip 1truecm
[J_{ab}, G_{c}]  &= -2 \delta_{c[a}G_{b]}\,,\nonumber\\[.2truecm]
[G_{a}, H]  &= - P_{a}\,, &
[G_{a},P_{b}]  &= -\delta_{ab}Z \,,\hskip 1.5truecm a=1,2.
\end{alignat}
For simpli\-ci\-ty, we consider the 3D case only. Without much change, the gauging procedure we describe below  also works in 4D. In table 1  we have indicated the symmetries, gauge fields, local parameters and curvatures that one associates to each of the generators.

{\small
\begin{table}[h]
\begin{center}
\begin{tabular}{|c|c|c|c|c|}
\hline
symmetry&generators& gauge field&parameters&curvatures\\[.1truecm]
\hline\rule[-1mm]{0mm}{6mm}
time translations&$H$&$\tau_\mu$&$\zeta(x^\nu)$&$R_{\mu\nu}(H)$\\[.1truecm]
space translations&$P^a$&$e_\mu{}^a$&$\zeta^a(x^\nu)$&$R_{\mu\nu}{}^a(P)$\\[.1truecm]
boosts&$G^a$&$\omega_\mu{}^{a}$&$\lambda^a(x^\nu)$&$R_{\mu\nu}{}^a(G)$\\[.1truecm]
spatial rotations&$J^{ab}$&$\omega_\mu{}^{ab}$&$\lambda^{ab}(x^\nu)$&$R_{\mu\nu}{}^{ab}(J)$\\[.1truecm]
central charge transf.&$Z$&$m_\mu$&$\sigma(x^\nu)$&$R_{\mu\nu}(Z)$\\[.1truecm]
\hline
\end{tabular}
\caption{This table indicates the generators of the Bargmann algebra and the  gauge fields, local parameters and curvatures that are associated to each of these generators.}
\end{center}
 \end{table}
 }

According to the Bargmann algebra \eqref{Bargmann} the gauge fields transform under  spatial rotations, boosts and central charge transformations as follows\footnote{All parameters in this section, as well as in section 4, are dependent on the coordinates $x^\mu$, even when not explicitly indicated. When we discuss the gauge fixing in later sections, we will always explicitly indicate the dependence on the time and/or space coordinates of the various parameters.}:
\begin{align} \label{bossymm1}
\delta \tau_\mu &=  0 \,, \nonumber \\[.1truecm]
\delta e_\mu{}^a &=  \lambda^a{}_b e_\mu{}^b + \lambda^a \tau_\mu \,, \nonumber \\[.1truecm]
\delta \omega_\mu{}^{ab} &=  \partial_\mu \lambda^{ab} + 2 \lambda^{c[a} \omega_\mu{}^{b]}{}_c \,, \\[.1truecm]
\delta \omega_\mu{}^{a} &=  \partial_\mu \lambda^{a} - \lambda^{b} \omega_\mu{}^a{}_b + \lambda^{ab} \omega_{\mu b} \,,\nonumber \\
\delta m_\mu &=  \partial_\mu \sigma + \lambda^a e_{\mu a}\nonumber\,.
\end{align}
The following curvatures transform covariantly under these transformations:
\begin{align}\label{curvatures}
R_{\mu \nu}(H) &=  2 \partial_{[\mu} \tau_{\nu]}\,, \nonumber
\\[.1truecm]
R_{\mu \nu}{}^a(P) &=  2 \partial_{[\mu} e_{\nu]}{}^a - 2
\omega_{[\mu}{}^{ab} e_{\nu]b} - 2
\omega_{[\mu}{}^a \tau_{\nu]}\,, \nonumber \\[.1truecm]
R_{\mu\nu}{}^a(G) &= 2\partial_{[\mu}\omega_{\nu]}{}^{a} - 2
\omega_{[\mu}{}^{ab} \omega_{\nu]b}\,,
\nonumber\\[.1truecm]
R_{\mu\nu}{}^{ab} (J) &=
2\partial_{[\mu}\omega_{\nu]}{}^{ab}  \,, \nonumber\\[.1truecm]
R_{\mu\nu}(Z) &= 2 \partial_{[\mu} m_{\nu]} - 2
\omega_{[\mu}{}^a e_{\nu]a}\,. \nonumber
\end{align}

We then proceed by imposing the following conventional constraints
\begin{equation}\label{bosoniccc}
R_{\mu\nu}{}^a(P)=0\,, \hskip 2truecm R_{\mu\nu}(Z)=0\,.
\end{equation}
On top of this, we impose the additional constraints:

\begin{equation}\label{bosonictc}
R_{\mu\nu}(H)=0\,,\hskip 2truecm R_{\mu\nu}{}^{ab}(J)=0\,.
\end{equation}
The first equation defines the foliation of a Newtonian spacetime. The second one is needed to obtain Newton gravity in flat space. The constraints \eqref{bosoniccc}, together with the first constraint of \eqref{bosonictc} can then be used to convert the $H$- and $P^a$-transformations, with parameters $\zeta(x^\nu)$ and $\zeta^a(x^\nu)$, of the algebra into
general coordinate transformations, with parameters $\xi^\lambda(x^\nu)$.

The gauge fields $\tau_\mu$ and $e_\mu{}^a$ can now be interpreted as the temporal and spatial Dreibeins.
Their projective inverses, $\tau^\mu$ and $e^\mu{}_a$, are defined as follows:
\begin{alignat}{2} \label{vielbrelSNC}
e_\mu{}^a e^\mu{}_b &= \delta^a_b \,, &\qquad \tau^\mu \tau_\mu = 1 \,, \nonumber \\[.2truecm]
\tau^\mu e_\mu{}^a &= 0 \,, & \tau_\mu e^\mu{}_a = 0 \,, \\[.2truecm]
e_\mu{}^a e^\nu{}_a &= \delta^\nu_\mu - \tau_\mu \tau^\nu \,. & \nonumber
\end{alignat}
Using these projective inverses one can use the conventional constraints \eqref{bosoniccc} to solve for the spin-connections fields
$\omega_\mu{}^{ab}(x^\nu)$ and $\omega_\mu{}^a(x^\nu)$ in terms of $\tau_\mu, e_\mu{}^a$ and $m_\mu$:
\begin{align}
 \omega_{\mu}{}^{ab}(x^\nu)
   & =   2 e^{\rho\,[a}   \partial_{[\rho}e_{\mu]}{}^{b]}
        + e_{\mu}{}^{c} e^{\rho\,a}e^{\nu\,b}  \partial_{[\rho}e_{\nu]}{}^{c}
        - \tau_{\mu} e^{\rho\,a}e^{\nu\,b}\partial_{[\rho} m_{\nu]}
           \,,  \label{NComega1}\\[.2truecm]
\omega_{\mu}{}^{a}(x^\nu)
   & = e^{\nu\, a} \partial_{[\mu} m_{\nu]}
               + e_{\mu}{}^b e^{\nu\, a}\tau^{\rho} \partial_{[\nu}e_{\rho]}{}^{b}
 + \tau^{\nu} \partial_{[\mu}e_{\nu]}{}^{a}  + \tau_{\mu} \tau^{\nu}e^{\rho\,a} \partial_{[\nu} m_{\rho]}\,.  \label{NComega2}
\end{align}

At this point, the only non-zero curvature left is the one corresponding to the boost transformations. Plugging the previous constraints into the Bianchi identities one finds that the only non-zero components of the boost curvature are given by
\begin{equation}
R_{0(a,b)}(G)\ne 0\,.
\end{equation}
The dynamical equation defining Newton-Cartan gravity is given by the trace of the above expression:
\begin{equation}\label{bosoniceom}
R_{0,a,a}(G)=0\,.
\end{equation}
These equations of motion are invariant under general coordinate transformations, local boosts, local
spatial rotations and local central charge transformations, with parameters $\xi^\lambda(x^\mu), \lambda^a(x^\mu),
\lambda^{ab}(x^\mu)$ and $\sigma(x^\mu)$, respectively.

\subsection{Galilean gravity}

To obtain Galilean gravity, described in terms of a Newton potential $\Phi(x^\mu)$, we perform a partial gauge fixing of the Newton-Cartan theory.
We now describe the details of this partial gauge fixing. First, we solve the constraints \eqref{bosonictc}
by imposing the gauge fixing conditions
\begin{equation}\label{gffirst}
\tau_\mu(x^\nu) = \delta_\mu{}^\emptyset\,,\hskip 1.5truecm \omega_\mu{}^{ab}(x^\nu)=0\,.
\end{equation}
This fixes the local time translations and spatial
rotations to constant transformations:
 \begin{equation}
 \xi^\emptyset(x^\nu) = \xi^\emptyset\,,\hskip 2truecm \lambda^{ab}(x^\nu)=\lambda^{ab}\,.
 \end{equation}
 No compensating transformations are induced by these gauge fixings.
Next, we gauge fix the spatial dependence of the spatial translations by imposing the gauge fixing condition
\begin{equation}\label{gffs}
e_i{}^a(x^\nu) = \delta_i{}^a\,.
\end{equation}
Requiring $\delta e_i{}^a=0$ leads to the condition
\begin{equation}\label{solgf}
\xi^a(x^\nu)=\xi^a(t) -\lambda_{ai}x^i\,.
\end{equation}
 The solution \eqref{solgf} for the spatial dependence of the spatial translation parameters
 expresses the fact that, after imposing the gauge fixing condition \eqref{gffs}, the $i$ index should be treated as an $a$ index and therefore only feels the constant spatial rotations.
 Note that after imposing the gauge fixing \eqref{gffs} space is flat and we do not distinguish anymore between the $i$ and $a$ indices and upper and down indices.

 At this stage the independent
 temporal and spatial Dreibein components and their projective inverses are given by\,\footnote{Remember that $\tau^i = \tau^a\delta^i_a$ and that we do not distinguish between $\tau^i$ and $\tau^a$ anymore.}
 \begin{alignat}{2}\label{FlatspatialcoordSNC}
\tau_\mu(x^\nu) &= \delta_{\mu}{}^\emptyset \,,&\hskip 2.5truecm  e_\mu{}^{a}(x^\nu) &= \bigl(-\tau^a(x^\nu)\,, \delta_i{}^a\bigr)\,,\nonumber\\[.15truecm]
\tau^\mu(x^\nu) &= \bigl(1\,, \tau^a(x^\nu) \bigr)\,,& e^{\mu}{}_{a}(x^\nu) &= \bigl(0\,, \delta^i{}_a\bigr)\,,
\end{alignat}
where the $\tau^a(x^\nu)$ are the only non-constant Dreibein components left. The only other independent gauge field left is the central charge gauge field $m_\mu(x^\nu)$. Ta\-king into account the compensating gauge transformation  given in
\eqref{solgf} we find that the remaining fields $\tau^a(x^\nu), m_\emptyset(x^\nu)$ and $m_i(x^\nu)$ transform as follows:
\begin{align}
\delta \tau^a(x^\nu) &= \lambda^a{}_b \tau^b(x^\nu) - \lambda^c{}_d x^d \partial_c \tau^a(x^\nu) + \xi^{\slashed 0} \partial_{\slashed 0} \tau^a(x^\nu) + \xi^j(t) \partial_j \tau^a(x^\nu)-\nonumber\\[.1truecm]
 &\ \ \ - \dot{\xi}^a(t) - \lambda^a (x^\nu) \,,\label{transftaui}\\[.2truecm]
\delta m_i(x^\nu) & =  \xi^{\slashed 0} \partial_{\slashed 0} m_i(x^\nu) + \xi^j(t) \partial_j m_i(x^\nu) + \lambda_i{}^j m_j(x^\nu) - \lambda^j{}_k x^k \partial_j m_i(x^\nu) +  \nonumber \\[.1truecm] &  \ \ \ +\lambda_i(x^\nu)
 + \partial_i \sigma(x^\nu)\,, \label{transfmi} \\[.2truecm]
\delta m_{\slashed 0}(x^\nu) & =  \xi^{\slashed 0} \partial_{\slashed 0} m_{\slashed 0}(x^\nu) + \dot{\xi}^i(t) m_i(x^\nu) + \xi^i(t) \partial_i m_{\slashed 0}(x^\nu) - \lambda^i{}_j x^j \partial_i m_{\slashed 0}(x^\nu) - \nonumber \\[.1truecm] &  \ \ \ -\lambda^a(x^\nu) \tau_a(x^\nu)+ \dot{\sigma}(x^\nu) \label{transfm0}\,.
\end{align}

The three fields $\tau^a(x^\nu)$, $m_i(x^\nu)$ and $m_{\slashed 0}(x^\nu)$ are not independent.
Since the gauge field $\omega_\mu{}^{ab}(x^\nu)$ which we gauge fixed to zero, see eq.~\eqref{gffirst}, is dependent we need to investigate its consequences. It turns out that the spatial part of these conditions does not lead to restrictions on the above fields but the time component does. Using the other gauge fixing conditions as well, we find that the
gauge fixing condition $\omega_\emptyset{}^{ab}(x^\nu)=0$ leads to the following restriction:
\begin{equation}\label{eq}
\partial_{[i}\tau_{j]}(x^\nu) + \partial_{[i}m_{j]}(x^\nu)=0\,.
\end{equation}
This implies that, locally,  one can write\,\footnote{Note that we freely lower and raise the $i$ or $a$ index on $\tau^i$ here and in the following. So, $\tau_i$ no longer refers to the $i$-components of $\tau_\mu$ at this point.}
\begin{equation}
\tau_i (x^\nu) + m_i(x^\nu) = \partial_i m(x^\nu) \,.
\end{equation}
Without loss of generality, we can thus eliminate $m_i(x^\nu)$ for $\tau_i(x^\nu)$ and $m(x^\nu)$, which is what we will do in the following.
The transformation rule for $m(x^\nu)$ can be found from $\delta \tau_i(x^\nu)$ and $\delta m_i(x^\nu)$:
\begin{equation} \label{transfm}
\delta m(x^\nu) = \xi^{\slashed 0} \partial_{\slashed 0} m(x^\nu) - \dot{\xi}^k(t) x^k + \xi^j(t) \partial_j m(x^\nu) - \lambda^j{}_k x^k \partial_j m(x^\nu) + \sigma(x^\nu) + Y(t) \,,
\end{equation}
where $Y(t)$ is an arbitrary time-dependent shift.
At this point we are left with three independent fields $\tau^i(x^\nu)$, $m_{\slashed 0}(x^\nu)$ and $m(x^\nu)$ whose transformation laws are given by \eqref{transftaui}, \eqref{transfm0}, \eqref{transfm}, respectively.

From the transformation rule \eqref{transfm}, we see that the central charge transformation acts as a St\"uckelberg shift on the field $m(x^\nu)$. We can thus partially fix the central charge transformations by imposing
\begin{equation} \label{centrcond}
m(x^\nu) = 0 \,.
\end{equation}
This fixes the central charge transformations according to
\begin{equation}\label{compcc}
\sigma (x^\mu) = \sigma(t) + \dot{\xi}^a(t)x_a\,,
\end{equation}
where it is understood that we also fix $Y(t)=-\sigma(t)$ in \eqref{transfm}.
After this gauge fixing the transformation rules of the two independent fields $\tau^i(x^\nu)$ and $m_{\slashed 0}(x^\nu)$ are given by:
\begin{align} \label{finallaws}
\delta \tau^i(x^\nu) &=  \lambda^i{}_j \tau^j(x^\nu) - \lambda^j{}_k x^k \partial_j \tau^i(x^\nu) + \xi^{\slashed 0} \partial_{\slashed 0} \tau^i(x^\nu) + \xi^j(t) \partial_j \tau^i(x^\nu) - \dot{\xi}^i(t) - \lambda^i (x^\nu) \,, \nonumber \\[.2truecm]
\delta m_{\slashed 0}(x^\nu) &=  \xi^{\slashed 0} \partial_{\slashed 0} m_{\slashed 0}(x^\nu) - \dot{\xi}^i(t) \tau_i(x^\nu) + \xi^i(t) \partial_i m_{\slashed 0}(x^\nu) + \ddot{\xi}^k (t) x^k \nonumber \\[.1truecm] & \ \ \  - \lambda^i{}_j x^j \partial_i m_{\slashed 0}(x^\nu) - \lambda^i(x^\nu) \tau_i(x^\nu)  + \dot{\sigma}(t) \,.
\end{align}
We note that the local boost transformations, with local parameters $\lambda^i(x^\nu)$,  end up as a Stueckelberg symmetry.
This Stueckelberg symmetry can be fixed by imposing the final gauge condition
\begin{equation}\label{gctau}
\tau^a(x^\nu)=0\,.
\end{equation}
This leads to the following compensating transformations:
\begin{equation}\label{comptau}
\lambda^i(x^\nu) =  - \dot{\xi}^i(t)\,.
\end{equation}
The only independent field left now is
\begin{equation}
 m_{\slashed 0}(x^\nu)\equiv \Phi(x^\nu)\,,
 \end{equation}
 which in a minute we will identify as the Newton potential.  Using the gauge condition \eqref{gctau} and taking into account the compensating transformations \eqref{comptau} we find that the transformation rule of this field is given by
\begin{equation}
\delta \Phi(x^\nu) = \xi^{\slashed 0} \partial_{\slashed 0} \Phi(x^\nu) + \xi^i(t) \partial_i \Phi (x^\nu)+ \ddot{\xi}^k(t) x^k - \lambda^i{}_j x^j \partial_i \Phi(x^\nu) + \dot{\sigma}(t) \,.
\label{gctPhiNC}
\end{equation}

The fact that we identify the field $m_{\slashed 0}(x^\nu)$ with the Newton potential $\Phi(x^\nu)$ is justified by looking at the
equations of motion.
In terms of  $\Phi(x^\nu)$
the expressions for the only non-zero dependent boost spin-connection field, see eq.~\eqref{NComega2}, is given by
\begin{equation}\label{boostcomp}
\omega_\emptyset{}^a(x^\nu) = - \partial^a\Phi(x^\nu)\,.
\end{equation}
If we now plug this expression for the boost spin-connection components into the
equation of motion \eqref{bosoniceom} we find the expected Poisson equation for the Newton potential:
\begin{equation}
\triangle\Phi = \partial_a\partial_a\Phi=0\,.
\end{equation}
This equation is invariant under the acceleration extended Galilei symmetries \eqref{gctPhiNC}.

The transformations \eqref{gctPhiNC}  close an algebra on $\Phi(x^\nu)$. 
One finds the following non-zero commutators:
\begin{align} \label{boscomm}
\left[ \delta_{\xi^{\slashed 0}}, \delta_{\xi^i(t)} \right] \Phi(x^\nu) & = \delta_{\xi^i(t)}\left(-\xi^{\slashed 0} \dot{\xi}^i(t)\right) \Phi(x^\nu) \,, \nonumber \\[.1truecm]
\left[ \delta_{\xi^{\slashed 0}}, \delta_{\sigma(t)} \right] \Phi(x^\nu) & =  \delta_{\sigma(t)} \left(- \xi^{\slashed 0} \dot{\sigma}(t) \right) \Phi(x^\nu) \,, \nonumber \\[.1truecm]
\left[ \delta_{\xi^i_1(t)}, \delta_{\xi^i_2(t)} \right] \Phi(x^\nu) & =  \delta_{\sigma(t)}\left(\dot{\xi}_1^j (t) \xi_2^j (t) - \dot{\xi}_2^j (t) \xi_1^j (t)\right) \Phi(x^\nu)\,, \nonumber \\[.1truecm]
\left[\delta_{\xi^i(t)} ,\delta_{\lambda^{jk}} \right] \Phi(x^\nu) & =  \delta_{\xi^i(t)}\left(\lambda^i{}_j \xi^j(t) \right) \Phi(x^\nu) \,,
\end{align}
where we have indicated the parameters of the transformations on the right-hand-side in the brackets.
Note that in calculating the commutator on $\Phi(x^\nu)$ we do not vary the explicit $x^a$ that occurs  in this transformation rule. This $x^a$-dependence follows from solving a parameter, see eq.~\eqref{compcc},  and we do not vary the parameters of the transformations when calculating commutators.

This finishes our review of the bosonic case. For the convenience of the reader we have summarized
all gauge conditions and resulting compensating transformations in Table 2.

{\small
\begin{table}[h]
\begin{center}
\begin{tabular}{|c|c|}
\hline
gauge condition/restriction&compensating transformation\\[.1truecm]
\hline\rule[-1mm]{0mm}{6mm}
$\tau_\mu(x^\nu)=\delta_\mu{}^\emptyset$&$\xi^\emptyset(x^\nu) =\xi^\emptyset$\\[.1truecm]
$\omega_\mu{}^{ab}(x^\nu)=0$&$\lambda^{a}(x^\nu)=\lambda^{ab}$\\[.1truecm]
$e_i{}^a(x^\nu)=\delta_i{}^a$&$\xi^a(x^\nu) = \xi^a(t)-\lambda_{ai}x^i$\\[.1truecm]
$\tau_i(x^\nu) + m_i(x^\nu) = \partial_i m(x^\nu)$& --\\[.1truecm]
$m(x^\nu)=0$&$\sigma(x^\nu) = \sigma(t)+{\dot{\xi}}^a(t) x_a$\\[.1truecm]
$\tau^a(x^\nu)=0$&$\lambda^i(x^\nu) = -{\dot \xi}^i(t)$\\[.1truecm]
\hline\hline
$m_\emptyset(x^\nu) = \Phi(x^\nu)$&$\omega_\emptyset{}^a(x^\nu) = -\partial^a\Phi(x^\nu)$\\[.1truecm]
\hline
\end{tabular}
\caption{This table indicates the gauge fixing conditions and corresponding compensating transformations that lead to Galilean gravity. We have also included the restrictions that follow from the fact that the spin-connection field $\omega_\mu{}^{ab}$ is dependent. At the bottom of the table we have summarized the expressions of the
non-zero remaining gauge fields in terms on the Newton potential $\Phi(x^\nu)$.}
\end{center}
 \end{table}
 }

\section{The 3D $\mathcal{N}=2$ Super-Bargmann Algebra}

A supersymmetric extension of the Bargmann algebra can be obtained
by contracting the super-Poincar\'{e} algebra with a central
extension, similar to how the Bargmann algebra can be obtained from
a trivially extended Poincar\'{e} algebra. It turns out that in
order to obtain a true supersymmetric extension of the
Bargmann algebra  in which the anti-commutator of two supersymmetry
generators gives both a time and a space  translation we need at
least two supersymmetries \cite{Gomis:2004pw}. In this work we will
consider the minimal case, i.e. $\mathcal{N}=2$ supersymmetry.

Our starting point is therefore the 3D $\mathcal{N}=2$ super-Poincar\'{e} algebra with central extension $\Z$, whose non-zero commutation relations are
given by
\begin{align}
[M_{BC},P_{A}] & = -2\eta_{A[B}P_{C]}\,, \ \ \ \ \hskip 1truecm [M_{CD},M_{EF}]  = 4 \eta_{[C [E}M_{F]D]} \,, \nonumber\\[.2truecm]
[M_{AB}, Q_{\alpha}] & = - \frac{1}{2}[\gamma_{AB}]_{\alpha}{}^{\beta} Q_{\beta}\,, \nonumber\\[.2truecm]
\{Q_{\alpha}^i \, ,\, Q_{\beta}^j \} & = -  \, [\gamma^A \gamma^{0}
]_{\alpha\beta} P_A\delta^{ij} +  \epsilon_{\alpha\beta}
\,\epsilon^{ij}\Z \,. \label{SuperPoincarealgebra}
\end{align}
The indices $A,B,\dots = 0,1,2$ are flat Lorentz indices,
$\alpha=1,2$ are 3D spinor indices and $i=1,2$ count the number of
supercharges. We have collected the 4 supercharges into two
2-component Majorana spinors $Q_\alpha^i$.\,\footnote{We use a
Majorana representation for the $\gamma$-matrices, in which the
charge conjugation matrix $C$ is given by $C = \gamma^0$.}

Following \cite{Lukierski:2006tr}, we define the linear combinations
\begin{equation}
Q_\alpha^\pm \equiv Q_\alpha^1 \pm \epsilon_{\alpha\beta}Q_\beta^2
\end{equation}
and apply the following rescaling, with a real parameter $\omega$,  of the generators and the central extension:
\begin{align} \label{SuperPoinRescalings}
Q^-_\alpha & \rightarrow \sqrt{\omega} Q^-_\alpha \,, \ \ \ \ \ \ \ \ \hskip 1truecm Q^+_\alpha \rightarrow \frac{1}{\sqrt{\omega}} Q^+_\alpha \,, \\
\Z & \rightarrow -\omega Z + \frac{1}{\omega}H \,, \ \ \ \hskip .5truecm  P_{0} \rightarrow \omega Z  + \frac{1}{\omega} H \,, \hskip 1truecm
M_{a0}  \rightarrow \omega G_{a}\,. \nonumber
\end{align}
We furthermore rename $M_{ab}=J_{ab}$.

The non-relativistic contraction  of the algebra \eqref{SuperPoincarealgebra}
is now defined by taking the limit $\omega \rightarrow \infty$.
This leads to the following 3D $\mathcal{N}=2$ super-Bargmann algebra:
\begin{alignat}{2}\label{N=2SuperBargmann}
[J_{ab},P_{c}]  &= -2\delta_{c[a}P_{b]}\,, & \qquad \qquad \qquad
[J_{ab}, G_{c}]  &= -2 \delta_{c[a}G_{b]}\,,\nonumber\\
[G_{a}, H]  &= - P_{a}\,, &
[G_{a},P_{b}]  &= -\delta_{ab}Z \,, \nonumber \\
[J_{ab}, Q^{\pm}] & = - \frac{1}{2}\gamma_{ab}Q^{\pm}  \,, &
[G_{a}, Q^+]  &= - \frac{1}{2}\gamma_{a0} Q^-  \,,\\
\{Q^{+}_\alpha \, ,\, Q^{+}_\beta \} & =  2\delta_{\alpha\beta} H
\,, & \{Q^{+}_\alpha \, ,\, Q^{-}_\beta \}  &= - \,
[\gamma^{a0}]_{\alpha\beta} P_{a} \,, \nonumber\\[.2truecm] \{Q^{-}_\alpha
\, ,\, Q^{-}_\beta \}  &=  2 \delta_{\alpha\beta}\, Z  \,.\nonumber
\end{alignat}
The bosonic part of the above algebra is the Bargmann algebra,
involving the Hamiltonian $H$, the spatial translations $P_a$, the
spatial rotations $J_{ab}$, the Galilean boosts $G_a$ and the
central charge $Z$.  Note that the bosonic Bargmann generators and
the central charge, together with the fermionic $Q^-$ generators
form the following $\mathcal{N}=1$ subalgebra\cite{Clark:1983ne}\,:
\begin{alignat}{2}\label{N=2SuperBargmannsub}
[J_{ab},P_{c}]  &= -2\delta_{c[a}P_{b]}\,, & \hskip 1truecm
[J_{ab}, G_{c}]  &= -2 \delta_{c[a}G_{b]}\,,\nonumber\\[.2truecm]
[G_{a}, H]  &= - P_{a}\,, &
[G_{a},P_{b}]  &= -\delta_{ab}Z \,, \\[.2truecm]
[J_{ab}, Q^{-}] & = - \frac{1}{2}\gamma_{ab}Q^{-} \,, & \{Q^{-}_\alpha \, ,\, Q^{-}_\beta \} &= 2
\delta_{\alpha\beta}\, Z \,.\nonumber
\end{alignat}
 The same does not apply if we
include the $Q^+$ generators instead of the $Q^-$ generators. This is due to the $[G,Q]$ commutator,
see \eqref{N=2SuperBargmann}, in which the $Q^+$ and $Q^-$
generators occur asymmetrically. The $\mathcal{N}=1$ sub-algebra
\eqref{N=2SuperBargmannsub} is not a true supersymmetry algebra in
the sense that the anti-commutator of two $Q^-$ supersymmetries does
not give a time and space translation but a central charge
transformation. Although the  $\mathcal{N}=2$ supersymmetry algebra
\eqref{N=2SuperBargmann} is a true supersymmetry algebra the
converse is not true: not every $\mathcal{N}=2$ super-algebra is
necessarily a true supersymmetry algebra.

Finally, we  note that the above 3D $\mathcal{N}=2$
super-Bargmann algebra can be embedded, via a null reduction, into a
$\mathcal{N}=1$ super-Poincar\'e algebra \cite{Kogut:1969xa}.

\section{3D $\mathcal{N}$=2 Newton-Cartan Supergravity}

In this section we apply a gauging procedure to the ${\cal
N}=2$ super-Bargmann algebra \eqref{N=2SuperBargmann} thereby extending the bosonic discussion of section 2 to the supersymmetric case.
As a first
step in  this gauging procedure we associate a gauge field to each
of the symmetries of the ${\cal
N}=2$ super-Bargmann algebra and we promote the constant parameters describing these
transformations  to arbitrary functions of the spacetime
coordinates $\{x^\mu\}$, see table 3.

{\small
\begin{table}[h]
\begin{center}
\begin{tabular}{|c|c|c|c|c|}
\hline
symmetry&generators& gauge field&parameters&curvatures\\[.1truecm]
\hline\rule[-1mm]{0mm}{6mm}
time translations&$H$&$\tau_\mu$&$\zeta(x^\nu)$&${\hat R}_{\mu\nu}(H)$\\[.1truecm]
space translations&$P^a$&$e_\mu{}^a$&$\zeta^a(x^\nu)$&${\hat R}_{\mu\nu}{}^a(P)$\\[.1truecm]
boosts&$G^a$&$\omega_\mu{}^{a}$&$\lambda^a(x^\nu)$&${\hat R}_{\mu\nu}{}^a(G)$\\[.1truecm]
spatial rotations&$J^{ab}$&$\omega_\mu{}^{ab}$&$\lambda^{ab}(x^\nu)$&${\hat R}_{\mu\nu}{}^{ab}(J)$\\[.1truecm]
central charge transf.&$Z$&$m_\mu$&$\sigma(x^\nu)$&${\hat R}_{\mu\nu}(Z)$\\[.1truecm]
\hline\hline
&&&&\\[-.45truecm]
two supersymmetries&$Q^\pm_\alpha$&$\psi_{\mu\pm}$&$\epsilon_\pm(x^\nu)$&${\hat\psi}_{\mu\nu\pm}$\\[.1truecm]
\hline
\end{tabular}
\caption{This table indicates the generators of the ${\cal N}=2$ super-Bargmann algebra and the  gauge fields, local parameters and super-covariant curvatures that are associated to each of these generators. The fermionic generators are indicated below the double horizontal line.}
\end{center}
 \end{table}

The corresponding gauge-invariant curvatures, see table 3,  are given by:
\begin{align}\label{curvatures}
\hat{R}_{\mu \nu}(H) & =  2 \partial_{[\mu} \tau_{\nu]} - \frac12 \bar{\psi}_{[\mu+} \gamma^0 \psi_{\nu]+} \,, \nonumber
\\[.1truecm]
\hat{R}_{\mu \nu}{}^a(P) & =  2 \partial_{[\mu} e_{\nu]}{}^a - 2
\omega_{[\mu}{}^{ab} e_{\nu]b} - 2
\omega_{[\mu}{}^a \tau_{\nu]} - \bar{\psi}_{[\mu+} \gamma^a \psi_{\nu]-} \,, \nonumber \\[.1truecm]
{\hat R}_{\mu\nu}{}^a(G) &= 2\partial_{[\mu}\omega_{\nu]}{}^{a} - 2
\omega_{[\mu}{}^{ab} \omega_{\nu]b}\,,
\nonumber\\[.1truecm]
{\hat R}_{\mu\nu}{}^{ab} (J) &=
2\partial_{[\mu}\omega_{\nu]}{}^{ab}  \,, \nonumber\\[.1truecm]
\hat{R}_{\mu\nu}(Z) & =  2 \partial_{[\mu} m_{\nu]} - 2
\omega_{[\mu}{}^a e_{\nu]a} -  \bar{\psi}_{[\mu-} \gamma^0
\psi_{\nu]-} \,, \nonumber \\[.1truecm]
\hat{\psi}_{\mu \nu +} & =  2
\partial_{[\mu} \psi_{\nu]+} - \frac{1}{2} \omega_{[\mu}{}^{ab}
\gamma_{ab} \psi_{\nu]+} \,, \nonumber \\[.1truecm]
\hat{\psi}_{\mu \nu -} & =  2 \partial_{[\mu} \psi_{\nu]-} -
\frac{1}{2} \omega_{[\mu}{}^{ab} \gamma_{ab} \psi_{\nu]-} +
\omega_{[\mu}{}^a \gamma_{a0} \psi_{\nu]+} \,.
\end{align}

According to the $\mathcal{N}=2$ super-Bargmann algebra
\eqref{N=2SuperBargmann}  the gauge fields given in
table 3 transform under spatial rotations, boosts
and central charge transformations as follows:
\begin{align} \label{bossymm}
\delta \tau_\mu & =  0 \,, \nonumber \\[.1truecm]
\delta e_\mu{}^a & =  \lambda^a{}_b e_\mu{}^b + \lambda^a \tau_\mu \,, \nonumber \\[.1truecm]
\delta \omega_\mu{}^{ab} & =  \partial_\mu \lambda^{ab} + 2 \lambda^{c[a} \omega_\mu{}^{b]}{}_c \,, \nonumber \\[.1truecm]
\delta \omega_\mu{}^{a} & =  \partial_\mu \lambda^{a} - \lambda^{b} \omega_\mu{}^a{}_b + \lambda^{ab} \omega_{\mu b} \,, \\
\delta m_\mu & =  \partial_\mu \sigma + \lambda^a e_{\mu a} \,, \nonumber \\[.1truecm]
\delta \psi_{\mu +} & =  \frac{1}{4} \lambda^{ab} \gamma_{ab}
\psi_{\mu +} \,, \hskip 1.5truecm \delta \psi_{\mu -}  = \frac{1}{4}
\lambda^{ab} \gamma_{ab} \psi_{\mu -} - \frac{1}{2} \lambda^{a}
\gamma_{a0} \psi_{\mu +} \,.\nonumber
\end{align}
We will discuss the other transformations of the $\mathcal{N}=2$
super-Bargmann algebra below.

The next step in the gauging procedure is to impose a set of
constraints on the curvatures. We first impose the following set of conventional constraints:
\begin{equation} \label{convconstr}
\hat{R}_{\mu \nu}{}^a(P) = 0 \,, \hskip 2truecm \hat{R}_{\mu \nu}(Z) = 0 \,.
\end{equation}
These conventional constraints can be used to solve
for the spin connections in terms of the other gauge fields as follows:\,\footnote{The projective inverses
$\tau^\mu$ and $e^\mu{}_a$ of $\tau_\mu$ and $e_\mu{}^a$ are defined
in eq.~\eqref{vielbrelSNC}.}
\begin{align}
 \omega_{\mu}{}^{ab}
   & =   2 e^{\rho\,[a}  \Bigl( \partial_{[\rho}e_{\mu]}{}^{b]} - \frac{1}{2} \bar{\psi}_{[\rho\, +} \gamma^{b]} \psi_{\mu]-}\Bigl)
        + e_{\mu}{}^{c} e^{\rho\,a}e^{\nu\,b} \Bigl( \partial_{[\rho}e_{\nu]}{}^{c} - \frac{1}{2}  \bar{\psi}_{[\rho\, +} \gamma^c
\psi_{\nu]-}  \Bigr) \nonumber\\
       & - \tau_{\mu} e^{\rho\,a}e^{\nu\,b}\Bigl( \partial_{[\rho} m_{\nu]}
        -  \frac{1}{2}\bar{\psi}_{[\rho\, -} \gamma^0 \psi_{\nu]-} \Bigr)   \,,  \label{SNComega1}\\[.3truecm]
\omega_{\mu}{}^{a}
   & = e^{\nu\, a}\Bigl( \partial_{[\mu} m_{\nu]}
        -  \frac{1}{2}\bar{\psi}_{[\mu\, -} \gamma^0 \psi_{\nu]-} \Bigr)
       + e_{\mu}{}^b e^{\nu\, a}\tau^{\rho} \Bigl( \partial_{[\nu}e_{\rho]}{}^{b} - \frac{1}{2}  \bar{\psi}_{[\nu\, +} \gamma^b
\psi_{\rho]-}  \Bigr) \nonumber\\
& + \tau^{\nu}\Bigl( \partial_{[\mu}e_{\nu]}{}^{a} - \frac{1}{2}  \bar{\psi}_{[\mu\, +} \gamma^a
\psi_{\nu]-}  \Bigr) + \tau_{\mu} \tau^{\nu}e^{\rho\,a}\Bigl( \partial_{[\nu} m_{\rho]}
        -  \frac{1}{2}\bar{\psi}_{[\nu\, -} \gamma^0 \psi_{\rho]-} \Bigr)\,.  \label{SNComega2}
\end{align}
On top of this we impose the following additional constraints:
\begin{equation}\label{ncsuper}
{\hat R}_{\mu\nu}(H)=0\,,\hskip 1.5truecm {\hat\psi}_{\mu\nu+}=0\,,\hskip 1.5truecm {\hat R}_{\mu\nu}{}^{ab}(J)=0\,.
\end{equation}
The first constraint defines a foliation of Newtonian spacetime. As we will see below the second constraint follows by
supersymmetry from the first constraint and, similarly, the third constraint follows from the second one.  This third constraint defines flat space Newton-Cartan supergravity. Note that, unlike in the bosonic case, this constraint is enforced upon us by supersymmetry. The constraints \eqref{convconstr}, together with the first constraint of \eqref{ncsuper} can be used to convert
the time and space
translations into general coordinate transformations,
with parameter $\xi^\mu(x^\nu)$.

The supersymmetry
variation of the conventional constraints does not lead to new
constraints as they are used to determine the supersymmetry
transformation rules of the now dependent gauge fields \eqref{SNComega1} and \eqref{SNComega2}. We find the following rules
for these gauge fields:\,\footnote{Recall that ${\hat\psi}_{ab} = e_a{}^\mu e_b{}^\nu{\hat\psi}_{\mu\nu}$.}
\begin{align} \label{susyvarspinconn}
\delta_Q \omega_\mu{}^{ab} & =   -\frac{1}{2} \bar{\epsilon}_+
\gamma^{[a} \hat{\psi}^{b]}{}_{\mu -} + \frac{1}{4} e_{\mu c}
\bar{\epsilon}_+ \gamma^c \hat{\psi}^{ab}{}_- - \frac{1}{2}
\tau_\mu \bar{\epsilon}_- \gamma^0 \hat{\psi}^{ab}{}_- \nonumber \\[.1truecm]
 & \ \ \ -\frac{1}{2} \bar{\epsilon}_-
\gamma^{[a} \hat{\psi}^{b]}{}_{\mu +} + \frac{1}{4} e_{\mu c}
\bar{\epsilon}_- \gamma^c \hat{\psi}^{ab}{}_+ \,, \nonumber
\\[.2truecm]
\delta_Q \omega_\mu{}^a & =  \frac{1}{2} \bar{\epsilon}_- \gamma^0
\hat{\psi}_\mu{}^a{}_- + \frac{1}{2} \tau_\mu \bar{\epsilon}_-
\gamma^0 \hat{\psi}_0{}^a{}_- + \frac{1}{4} e_{\mu b}
\bar{\epsilon}_+ \gamma^{b} \hat{\psi}^{a}{}_{0 -} + \frac{1}{4}
\bar{\epsilon}_+ \gamma^a \hat{\psi}_{\mu 0 -} \nonumber \\[.1truecm]
 & \ \ \  +
\frac{1}{4} e_{\mu b} \bar{\epsilon}_- \gamma^{b}
\hat{\psi}^{a}{}_{0 +} + \frac{1}{4} \bar{\epsilon}_- \gamma^a
\hat{\psi}_{\mu 0 +} \,.
\end{align}
In contrast,
we must investigate the supersymmetry variations of the
non-conventional constraints \eqref{ncsuper}. In order to do this, we must first determine the supersymmetry
rules of the independent gauge fields.

According to the super-Bargmann algebra \eqref{N=2SuperBargmann} the
supersymmetry transformations of the independent gauge fields are
given by
\begin{align} \label{ansatzsusyvar}
\delta_Q \tau_\mu & =  \frac{1}{2} \bar{\epsilon}_+ \gamma^0
\psi_{\mu +} \,, \nonumber
\\[.1truecm]
\delta_Q e_\mu{}^a & =  \frac{1}{2} \bar{\epsilon}_+ \gamma^a \psi_{\mu -} +  \frac{1}{2} \bar{\epsilon}_- \gamma^a \psi_{\mu +}  \,,  \nonumber \\
\delta_Q m_\mu & =  \bar{\epsilon}_- \gamma^0 \psi_{\mu -}  \,,
\nonumber
\\[.1truecm]
\delta_Q \psi_{\mu +} & =  D_\mu \epsilon_+  \,, \nonumber
\\[.1truecm]
\delta_Q \psi_{\mu -} & =  D_\mu \epsilon_- + \frac{1}{2}
\omega_\mu{}^{a} \gamma_{a0}  \epsilon_+  \,,
\end{align}
where $\omega_\mu{}^a$ is the dependent boost gauge field. The
covariant derivative $D_\mu$ is only covariantized with respect to
spatial rotations. When acting on the parameters $\epsilon_\pm$, it
is given by
\begin{equation}
D_\mu \epsilon_\pm = \partial_\mu \epsilon_\pm - \frac{1}{4}
\omega_\mu{}^{ab} \gamma_{ab} \epsilon_\pm
\end{equation}
in terms of the dependent connection field $\omega_\mu{}^{ab}$.

At this point we have  obtained the supersymmetry rules of {\sl all} gauge fields, both the dependent as well as the independent ones. We find that with these supersymmetry transformations the
supersymmetry algebra closes on-shell. To be precise, the commutator
of two supersymmetry transformations  closes and is given by the
following soft algebra:
\begin{align} \label{closbos}
\left[\delta_Q(\epsilon_1),\delta_Q(\epsilon_2)\right] & =
\delta_{\mathrm{g.c.t.}}(\xi^{\mu}) + \delta_{J_{ab}}(\lambda^a{}_b)
+ \delta_{G_a}(\lambda^a{}) + \delta_{Q_+}(\epsilon_+)\ +
 \nonumber \\[.1truecm] & \ \ \ +\ \delta_{Q_-}(\epsilon_-) + \delta_{Z}(\sigma) \,,
\end{align}
provided the following equations hold:
\begin{equation}\label{eom}
\gamma^\mu \tau^\nu \hat{\psi}_{\mu \nu-} = 0 \,, \hskip 2truecm e^\mu{}_a
e^\nu{}_b \hat{\psi}_{\mu \nu-} = 0 \,.
\end{equation}
The first equation can be seen as an equation of motion, the second
one does not contain any time derivatives and should be viewed as a fermionic constraint.
Here $\mathrm{g.c.t.}$ denotes a general
coordinate transformation and the field-dependent parameters are
given by
\begin{align} \label{parclos}
\xi^\mu & =  \frac{1}{2} \Big(\bar{\epsilon}_{2+} \gamma^0
\epsilon_{1+}  \Big) \tau^\mu + \frac{1}{2} \Big(\bar{\epsilon}_{2+}
\gamma^a \epsilon_{1-} + \bar{\epsilon}_{2-} \gamma^a \epsilon_{1+}
\Big) e^\mu{}_a \,, \nonumber
\\[.1truecm]
\lambda^a{}_b & =  -\xi^\mu \omega_\mu{}^a{}_b  \,, \nonumber
\\[.1truecm]
\lambda^a & =  -\xi^\mu \omega_\mu{}^a  \,, \nonumber \\[.1truecm]
\epsilon_\pm & =  - \xi^\mu \psi_{\mu \pm} \,, \nonumber
\\[.1truecm]
\sigma & =  - \xi^\mu m_\mu + \left( \bar{\epsilon}_{2-} \gamma^0
\epsilon_{1-}  \right) \,.
\end{align}

We are now in a position  to investigate the supersymmetry
variations of the three constraints \eqref{ncsuper}
 and of the equation of motion/constraint  \eqref{eom}.
One may verify that under  supersymmetry the  first constraint in \eqref{ncsuper} transforms
to the second one and that the  supersymmetry variation of the second constraint leads to the third one.
This third constraint
does not lead to new constraints  because the
supersymmetry variation of $\omega_\mu{}^{ab}$ vanishes on-shell, see
eq.~\eqref{susyvarspinconn}.

Substituting the constraints into the super-Bianchi identities, it follows that the only non-zero bosonic curvature we are left with is the boost
curvature ${\hat R}_{\mu\nu}{}^a(G)$ and we find that  only the following components are non-vanishing:
\begin{equation}
\tau^\mu e^\nu{}_{(a} {\hat
R}_{\mu\nu}{}^{b)}(G)\equiv {\hat
R}_{0(a}{}^{b)}(G)\ne 0\,.
\end{equation}
Using this it follows  that the supersymmetry variation of the second constraint in \eqref{eom}
does not lead to a new constraint. On the other hand, the supersymmetry variation of the fermionic equation of motion, i.e.~the first constraint  in \eqref{eom},
leads to the bosonic equation of motion
\begin{equation}\label{sbosoniceom}
{\hat R}_{0a}{}^{a}(G)=0\,.
\end{equation}

To finish the consistency check of the gauging  procedure we should check whether the supersymmetry variation of
 the bosonic equation of motion
\eqref{sbosoniceom} does not lead to new constraints and/or equations of motion. Instead of doing this we shall show
in the next section that after gauge fixing all constraints can be solved leading to a consistent system with a closed algebra.

This finishes our construction  of the 3D
$\mathcal{N}=2$ Newton-Cartan supergravity theory.

\section{3D Galilean Supergravity}

In this section we will perform a partial gauge fixing of the bosonic and fermionic symmetries to
derive the Newton-Cartan supergravity theory from the Galilean observer point of view. We will define a supersymmetric Galilean observer as one for which only a supersymmetric
extension of the acceleration extended Galilei symmetries are retained. Due to the constant
time translations, this implies in particular that only half of the supersymmetries will be gauged, see below.
 We closely follow the analysis given in section 2 for the bosonic case.

First, we solve the constraints \eqref{ncsuper}  by imposing the gauge fixing conditions
\begin{equation}\label{gffirstsuper}
\tau_\mu(x^\nu) = \delta_\mu{}^\emptyset\,,\hskip 1truecm \omega_\mu{}^{ab}(x^\nu)=0\,, \hskip 1truecm \psi_{\mu +}(x^\nu)=0\,.
\end{equation}
This fixes the local time translations, spatial rotations and $\epsilon_+$ transformations to constant transformations:
 \begin{equation}
 \xi^\emptyset(x^\nu) = \xi^\emptyset\,,\hskip 1truecm \lambda^{ab}(x^\nu)=\lambda^{ab}\,,\hskip 1truecm \epsilon_+(x^\nu) = \epsilon_+\,.
 \end{equation}
 No compensating transformations are induced by these gauge fixings.

We now  partially gauge fix the spatial translations by imposing the gauge choice
\begin{equation}\label{gf11}
e_i{}^a(x^\nu)=\delta_i{}^a\,.
\end{equation}
This gauge choice implies that we may use from now on the expressions \eqref{FlatspatialcoordSNC}
for the temporal and spatial Dreibein components and their projective inverses. We will derive the required compensating transformation below.
First, using the above gauge choices and the fact that the
purely spatial components ${\hat R}_{ij}{}^a(G)$ of the curvatures of boost transformations
and the purely spatial components  ${\hat\psi}_{ij -}$ of the curvature of $\epsilon_-$ transformations are zero, for their expressions see eq.~\eqref{curvatures},
 we derive that
\begin{equation}
\partial_{[i}\omega_{j]}{}^a=0\,,\hskip 2truecm \partial_{[i}\psi_{j]-}=0\,.
\end{equation}
The first equation we solve locally by writing
 \begin{equation}
 \omega_i{}^a = \partial_i\omega^a\,,
 \end{equation}
where $\omega^a$ is a dependent field since $\omega_i{}^a$ is dependent. This also explains why we have not added a purely time-dependent
piece to the r.h.s.~of the above solution.

 We next partially gauge fix the $\epsilon_-$ transformations by imposing the gauge choice
\begin{equation}\label{gcgravitino}
\psi_{i-}(x^\nu)=0\,.
\end{equation}
This fixes the $\epsilon_-$ transformations according to
\begin{equation}
\epsilon_-(x^\nu) = \epsilon_-(t) - \frac12 \omega^a \gamma_{a0} \epsilon_+ \,.
\end{equation}
Given the gauge choice \eqref{gcgravitino} the spatial translations are now fixed without
the need for any fermionic compensating transformation. Indeed, from the total variation of the gauge fixing condition \eqref{gf11} we find:
\begin{equation}
\xi^i (x^\nu) = \xi^i(t) - \lambda^i{}_j x^j \,.
\end{equation}

At this point, we are left with the remaining fields $\tau^a$, $m_i$, $m_{\slashed 0}$ and $\psi_{\slashed 0 -}$. These fields are
not independent since the gauge field $\omega_\mu{}^{ab}$ which we gauge fixed to zero is dependent,
 see eq.~\eqref{SNComega1}. Like in the bosonic case, only the time
component $\omega_{\slashed 0}{}^{ab}=0$ leads to a restriction:\,\footnote{Recall that $\tau_i = \tau^a\delta_{ia}$. Note also that under supersymmetry the  variation of this constraint gives
$\bar{\epsilon}_+ \gamma_{[i} \partial_{j]} \psi_{\slashed 0-} = 0$
which is equivalent to the fermionic  equation of motion (which after gauge fixing takes the form \eqref{feom}). Therefore, this constraint is consistent with supersymmetry.}
\begin{equation}
\partial_{[i} \left(\tau_{j]} + m_{j]} \right)(x^\nu) = 0 \,.
\end{equation}
As in the bosonic case, this implies that we can write locally:
\begin{equation} \label{defm}
\tau_i(x^\nu) + m_i(x^\nu) = \partial_i m (x^\nu)\,.
\end{equation}
Without loss of generality we will use this equation to eliminate $m_i$ in terms of the other two fields. The variation of $m$ is determined by writing the variation of $\tau_i + m_i$ as a $\partial_i$-derivative. This is trivial for most of the terms, except for the $\epsilon_+$ term. Before addressing this issue below, it is convenient to write down the total variation of $\partial_i m$ instead of $m$. From eq.~\eqref{defm} we find
\begin{equation}
\delta \partial_i m = \xi^{\slashed 0} \partial_{\slashed 0} \partial_i m + \xi^j(t) \partial_j \partial_i m
 +\lambda_i{}^j\partial_j m - \lambda^m{}_n x^n \partial_m \partial_i m + \partial_i \sigma(x^\nu) - \dot{\xi}^i(t)  - \frac12 {\bar\epsilon}_+\gamma_i  \psi_{\slashed 0-} \,.
\end{equation}
Note that the terms proportional to the local boost parameters $\lambda^i(x^\nu)$ have cancelled out. We may now partially gauge fix the central charge transformations by putting
\begin{equation}
m(x^\nu)=0\,.
\end{equation}
We thus obtain
\begin{equation}
\partial_i\sigma(x^\nu) = {\dot\xi}^i(t) +
\frac12 {\bar\epsilon}_+\gamma_i  \psi_{\slashed 0-}(x^\nu) \,,
\end{equation}
which is sufficient to calculate the transformation rule of $\partial_i m_{\slashed 0}$. After this gauge fixing, taking into account all the compensating transformations, see table 4 below, and the restriction \eqref{defm} with $m=0$ substituted, we find the following transformation rules for the remaining independent fields:
\begin{align}
\delta \tau_i & =  \xi^{\slashed 0} \partial_{\slashed 0} \tau_i + \xi^j(t) \partial_j \tau_i - \dot{\xi}^i(t) + \lambda_{ij} \tau^j - \lambda^k{}_l x^l \partial_k \tau_i - \lambda_i (x^\nu) - \frac12 \bar{\epsilon}_+ \gamma_i \psi_{\slashed 0 -} \,, \nonumber \\[.2truecm]
\delta \partial_i m_{\slashed 0} & =  \xi^{\slashed 0} \partial_{\slashed 0}\partial_i m_{\slashed 0} + \xi^j(t) \partial_j \partial_i m_{\slashed 0} + \ddot{\xi}^i(t)  - \dot{\xi}^j(t)\partial_i \tau_j +\lambda_i{}^j\partial_j m_{\slashed 0}  - \lambda^m{}_n x^n \partial_m \partial_i m_{\slashed 0} -
\nonumber\\[.1truecm]
& \ \ \ - \partial_i\big(\lambda^j(x^\nu) \tau_j\big)   + \bar{\epsilon}_- (t) \gamma^0\partial_i \psi_{\slashed 0 -} + \frac12\partial_i\big( \omega^a \bar{\epsilon}_+ \gamma_a \psi_{\slashed 0-}\big) + \frac12 {\bar\epsilon}_+ \gamma_i  {\dot \psi}_{\slashed 0-} \,,  \\[0truecm]
\delta \psi_{\slashed 0-} & =  \xi^{\slashed 0} \partial_{\slashed 0} \psi_{\slashed 0-} + \xi^i(t) \partial_i \psi_{\slashed 0 -} - \lambda^i{}_j x^j \partial_i \psi_{\slashed 0-} + \frac14 \lambda^{ab} \gamma_{ab} \psi_{\slashed 0 -}  \nonumber \\[-.1truecm] & \ \ \  + \dot{\epsilon}_-(t) + \frac12 \left(\omega_{\slashed 0}{}^a - \dot{\omega}^a \right) \gamma_{a0} \epsilon_+ \,.\nonumber
\end{align}
Note that $\omega_{\slashed 0}{}^a$ and $\omega^a$  depend on the fields $\tau_i$, $m_{\slashed 0}$.
Using expression \eqref{SNComega2} for the dependent boost gauge field $\omega_\mu{}^a$ one can calculate that
\begin{align}
\omega_i{}^a &\equiv \partial_i\omega^a = -\partial_i\tau^a \hskip .5truecm \rightarrow\hskip .5truecm \omega^a = -\tau^a\,,\\[.1truecm]
\omega_{\slashed 0}{}^a &= -\dot{\tau}^a - \partial_a \left(m_{\slashed 0} - \frac12 \tau^i \tau^i \right)\,.
\end{align}

As a final step we now  fix the local boost transformations by imposing
\begin{equation}
\tau^i(x^\nu)=0\,,
\end{equation}
which leads to the following compensating transformations:
\begin{equation}
\lambda^i(x^\nu) = -{\dot\xi}^i(t) -\frac12 \bar{\epsilon}_+ \gamma_i \psi_{\slashed 0 -}(x^\nu) \,.
\end{equation}
One now finds that
\begin{equation}
\omega^a=0\,,\hskip 2truecm \omega_{\slashed 0}{}^a = -\partial^a m_{\slashed 0} \equiv - \partial^a\Phi\,,
\end{equation}
where $\Phi$ is the Newton potential. In
terms of the `Newton force' $\Phi_i$ and its supersymmetric partner $\Psi$ defined  by
\begin{equation}
\Phi_i = \partial_i\Phi\,,\hskip 2truecm \Psi = \psi_{\slashed 0 -}\,,
\end{equation}
one thus obtains the following transformation rules:
\begin{align}
\delta  \Phi_i &=  \xi^{\slashed 0} \partial_{\slashed 0} \Phi_i + \xi^j(t) \partial_j \Phi_i + \ddot{\xi}^i(t) + \lambda_i{}^j  \Phi_j- \lambda^m{}_n x^n \partial_m \Phi_i  + \bar{\epsilon}_-(t) \gamma^0 \partial_i \Psi +  \frac12 \bar{\epsilon}_+ \gamma_i \dot{\Psi}\label{transfforce} \,,\\[.2truecm]
\delta \Psi &= \xi^{\slashed 0} \partial_{\slashed 0} \Psi + \xi^i(t) \partial_i \Psi - \lambda^i{}_j x^j \partial_i \Psi + \frac14 \lambda^{ab} \gamma_{ab} \Psi   + \dot{\epsilon}_-(t) - \frac12 \Phi^i  \gamma_{i0} \epsilon_+ \label{transfPsi}\,.
\end{align}
Note that the central charge transformations only act on the Newton potential, not on the Newton force. Determining the transformation rule of the Newton potential $\Phi$ is non-trivial, due to the fact that the last term of \eqref{transfforce} cannot be manifestly written as a $\partial_i$-derivative.
The above transformation rules are consistent with the integrability condition
\begin{equation}\label{integrability}
\partial_{[i} \Phi_{j]}(x^\nu) = 0\,,
\end{equation}
 by virtue of the fermionic equations of motion \eqref{eom} which, after gauge fixing, take on the form
 \begin{equation}\label{feom}
\gamma^i\partial_i\Psi(x^\nu)=0\hskip .5truecm \Leftrightarrow\hskip .5truecm  \partial_{[i}\gamma_{j]}\Psi(x^\nu)=0\,.
 \end{equation}
Under supersymmetry these fermionic equations of motion lead to the following bosonic equation of motion:
\begin{equation}
\partial^i\Phi_i(x^\nu)=0\,.
\end{equation}
The same bosonic equation of motion also follows from eq.~\eqref{sbosoniceom} after gauge fixing.

In order to obtain transformation rules for the Newton potential $\Phi$ and its fermionic superpartner, we need to solve the fermionic equations of motion/constraint \eqref{feom}. The second form of this constraint makes it clear that the equations of motion are solved by a spinor $\chi$, that obeys:
\begin{equation} \label{defchi}
\gamma_i \Psi = \partial_i \chi \,.
\end{equation}
Note that this only determines $\chi$ up to a purely time-dependent shift. From \eqref{defchi}, it follows that $\chi$ obeys the constraint:
\begin{equation} \label{constrchi}
\gamma^1 \partial_1 \chi = \gamma^2 \partial_2 \chi \,.
\end{equation}
$\Psi$ can thus be expressed in terms of $\chi$ in a number of equivalent ways:
\begin{equation}
\Psi = \gamma^1 \partial_1 \chi = \gamma^2 \partial_2 \chi = \frac12 \gamma^i \partial_i \chi \,.
\end{equation}
It is now possible to determine the transformation rule of $\Phi$ by rewriting $\delta \Phi_i$ as a $\partial_i$-derivative:
\begin{equation}
\delta \Phi_i = \partial_i (\delta \Phi) \,.
\end{equation}
The resulting transformation rule for the Newton potential is
\begin{equation}
\delta \Phi = \xi^{\slashed 0} \partial_{\slashed 0} \Phi + \xi^i(t) \partial_i \Phi + \ddot{\xi}^i(t) x^i - \lambda^m{}_n x^n \partial_m \Phi + \frac12 \bar{\epsilon}_-(t) \gamma^{0i} \partial_i \chi + \frac12 \bar{\epsilon}_+ \dot{\chi} + \sigma(t) \,.
\end{equation}
Note that we have allowed for an arbitrary time-dependent shift $\sigma(t)$ in the transformation rule, whose origin stems from the fact that $\Phi_i = \partial_i \Phi$ only determines $\Phi$ up to an arbitrary time-dependent shift.

In order to determine the transformation rule of $\chi$, we try to rewrite $\gamma_i \delta \Psi$ as a $\partial_i$-derivative\footnote{Note that even though $\Psi = \frac12 \gamma^i \partial_i \chi$, the correct transformation rule of $\chi$ cannot be found by writing $\delta \Psi$ as $\frac12 \gamma^i \partial_i$ of an expression. In particular, one would miss the term involving the dual Newton potential $\Xi$ in the transformation rule of $\chi$. This is due to the fact that $\Psi = \frac12 \gamma^i \partial_i \chi$ is a consequence of the defining equations $\gamma_i \Psi = \partial_i \chi$, but is not equivalent to it.}:
\begin{equation}
\gamma_i \delta \Psi = \partial_i (\delta \chi) \,.
\end{equation}
Most of the terms in $\gamma_i \delta \Psi$ can be straightforwardly written as a $\partial_i$-derivative. Only for the $\epsilon_+$ transformation, the argument is a bit subtle. We thus focus on the terms in $\gamma_i \delta \Psi$, given by
\begin{equation}
-\frac12 \gamma_i \Phi^j \gamma_{j0} \epsilon_+ = -\frac12 \gamma_i \partial^j \Phi \gamma_{j0}  \epsilon_+ = -\frac12 \partial^j \Phi \gamma_{ij0} \epsilon_+ - \frac12 \partial_i \Phi \gamma_0 \epsilon_+ \,.
\end{equation}
The last term is already in the desired form. To rewrite the first term in the proper form, we note that the Newton potential $\Phi$ can be dualized to a `dual Newton potential' $\Xi$ via
\begin{equation} \label{defxi}
\partial_i \Phi = \varepsilon_{ij} \partial^j \Xi \,, \qquad \partial_i \Xi = - \varepsilon_{ij} \partial^j \Phi\,.
\end{equation}
Using the convention that $\gamma_{ij0} = \epsilon_{0ij} = \epsilon_{ij}$, we then get
\begin{equation}
-\frac12 \gamma_i \Phi^j \gamma_{j0} \epsilon_+ =
\frac12 \partial_i \Xi \epsilon_+ - \frac12 \partial_i \Phi \gamma_0 \epsilon_+ \,.
\end{equation}
One thus obtains the following transformation rule for $\chi$, which includes the dual Newton potential $\Xi$:
\begin{equation} \label{rulechi}
\delta \chi = \xi^{\slashed 0} \partial_{\slashed 0} \chi +  \xi^i(t) \partial_i \chi - \lambda^m{}_n x^n \partial_m \chi + \frac14 \lambda^{mn} \gamma_{mn} \chi + x^i \gamma_i\dot{\epsilon}_-(t) + \frac12 \Xi \epsilon_+ - \frac12 \Phi \gamma_0 \epsilon_+ + \eta(t)\,.
\end{equation}
Note that we have again allowed for a purely time-dependent shift $\eta(t)$, whose origin lies in the fact that \eqref{defchi} only determines $\chi$ up to a purely time-dependent shift.

In order to calculate the algebra on $\Phi$, $\chi$, we also need the transformation rule of the dual potential $\Xi$. This rule is determined by dualizing the transformation rule of $\Phi$:
\begin{equation}
\partial_i(\delta \Xi) = -\varepsilon_{ij} \partial^j (\delta \Phi) \,.
\end{equation}
By repeatedly using \eqref{defchi} and \eqref{defxi}, we get:
\begin{equation}
\delta \Xi = \xi^{\slashed 0} \partial_{\slashed 0} \Xi + \xi^i(t) \partial_i \Xi + \ddot{\xi}^i(t) \varepsilon_{ij} x^j - \lambda^m{}_n x^n \partial_m \Xi + \frac12 \bar{\epsilon}_-(t) \gamma^i \partial_i \chi - \frac12 \bar{\epsilon}_+ \gamma_0 \dot{\chi} + \tau(t) \,,
\end{equation}
where we again allowed for a purely time-dependent shift $\tau(t)$.

The algebra then closes on $\Phi$ and $\chi$, using \eqref{defchi}, \eqref{constrchi}, \eqref{defxi} . One finds the following non-zero commutators between the fermionic symmetries:
\begin{align}
\left[\delta_{\epsilon_{1-}(t)},\delta_{\epsilon_{2-}(t)} \right] &= \delta_{\sigma(t)}\left(\frac{\rmd}{\rmd t} \left(\bar{\epsilon}_{2-}(t) \gamma^0 \epsilon_{1-}(t) \right)\right) \,,\nonumber \\
\left[\delta_{\epsilon_{1+}},\delta_{\epsilon_{2+}} \right] &= \delta_{\xi^{\slashed 0}} \left(\frac12 \left(\bar{\epsilon}_{2+} \gamma^0 \epsilon_{1+}  \right) \right) \,, \nonumber \\
\left[\delta_{\epsilon_{+}},\delta_{\epsilon_{-}(t)} \right] &= \delta_{\xi^i(t)} \left( \frac12 \left(\bar{\epsilon}_-(t) \gamma^i \epsilon_+ \right)\right) \,, \nonumber \\
\left[\delta_{\eta(t)}, \delta_{\epsilon_+} \right] &= \delta_{\sigma(t)}\left(\frac12\left( \bar{\epsilon}_+ \dot{\eta}(t) \right) \right) \,.
\end{align}
The non-zero commutators between the bosonic and fermionic symmetries are given by:
\begin{alignat}{2}
\left[\delta_{\xi^i(t)},\delta_{\epsilon_{+}} \right] &= \delta_{\epsilon_{-}(t)} \left(\frac12 \dot{\xi}^i(t) \gamma_{0i} \epsilon_+ \right)\,, &\qquad \left[\delta_{\lambda^{ij}},\delta_{\epsilon_{+}} \right] &= \delta_{\epsilon_+} \left( -\frac14 \lambda^{ij} \gamma_{ij} \epsilon_+ \right) \,, \nonumber \\
\left[\delta_{\xi^{\slashed 0}},\delta_{\epsilon_{-}(t)} \right] &= \delta_{\epsilon_{-}(t)} \left(- \xi^{\slashed 0} \dot{\epsilon}_-(t) \right)\,, & \left[\delta_{\xi^i(t)},\delta_{\epsilon_{-}(t)} \right] &= \delta_{\eta(t)} \left( -\xi^i(t) \gamma_i \dot{\epsilon}_-(t) \right)\,, \nonumber \\
\left[\delta_{\lambda^{ij}},\delta_{\epsilon_{-}(t)} \right] &= \delta_{\epsilon_{-}(t)} \left( -\frac14 \lambda^{ij} \gamma_{ij} \epsilon_-(t) \right)\,, & \left[\delta_{\sigma(t)},\delta_{\epsilon_{+}} \right] &= \delta_{\eta(t)} \left(\frac12\left(\sigma(t) \gamma^0 \epsilon_+ \right) \right) \,, \nonumber \\
\left[\delta_{\xi^{\slashed 0}},\delta_{\eta(t)} \right] &= \delta_{\eta(t)} \left(-\xi^{\slashed 0} \dot{\eta}(t) \right)\,, & \left[\delta_{\lambda^{ij}},\delta_{\eta(t)} \right] &= \delta_{\eta(t)} \left( -\frac14 \lambda^{ij} \gamma_{ij} \eta(t) \right) \,.
\end{alignat}
The bosonic commutators are not changed with respect to the purely bosonic case and are given by \eqref{boscomm}.

It is interesting to comment on the appearance of holomorphic functions in the above description. In a basis in which
\begin{equation}
\gamma^1 = \left( \begin{array}{cc} 0 & 1 \\ 1 & 0 \end{array} \right) \,, \qquad \gamma^2 = \left( \begin{array}{cc} 1 & 0 \\ 0 & -1 \end{array} \right) \,,
\end{equation}
the constraint \eqref{constrchi} on $\chi$ reduces to the Cauchy-Riemann equations for a holomorphic function $\chi_2 + \rmi \, \chi_1$, where the indices 1, 2 refer to spinor indices. Interestingly, the appearance of the dual potential implies that a holomorphic function, given by $\Phi + \rmi \, \Xi$, also emerges in the bosonic sector. Indeed, the definition of \eqref{defxi} corresponds to the Cauchy-Riemann equations for this function. Both the real and imaginary parts of this holomorphic function then satisfy the two-dimensional Laplace equation.

This finishes our discussion of the ${\cal N}=2$ Galilean supergravity theory. Like in the bosonic case, see the end of section 2, we have summarized all gauge fixing conditions and resulting compensating transformations in table 4.

{\small
\begin{table}[h]
\begin{center}
\begin{tabular}{|c|c|}
\hline
gauge condition/restriction&compensating transformation\\[.1truecm]
\hline\rule[-1mm]{0mm}{6mm}
$\tau_\mu(x^\nu)=\delta_\mu{}^\emptyset$&$\xi^\emptyset(x^\nu) =\xi^\emptyset$\\[.1truecm]
$\omega_\mu{}^{ab}(x^\nu)=0$&$\lambda^{a}(x^\nu)=\lambda^{ab}$\\[.1truecm]
$\psi_{\mu +}(x^\nu)=0$&$\epsilon_+(x^\nu)=\epsilon_+$\\[.1truecm]
$e_i{}^a(x^\nu)=\delta_i{}^a$&$\xi^i(x^\nu) = \xi^i(t)-\lambda^i{}_jx^j$\\[.1truecm]
$\psi_{i-}(x^\nu)=0$&$\epsilon_-(x^\nu) = \epsilon_-(t) -\tfrac{1}{2}\omega^a(x^\nu)\gamma_{a0}\epsilon_+$\\[.1truecm]
$\tau_i(x^\nu) + m_i(x^\nu) = \partial_i m(x^\nu)$& --\\[.1truecm]
$m(x^\nu)=0$&$\partial_i\sigma(x^\nu) = {\dot\xi}^i(t) +
\frac12 {\bar\epsilon}_+\gamma_i  \psi_{\slashed 0-}(x^\nu)$\\[.1truecm]
$\tau^a(x^\nu)=0$&$\lambda^i(x^\nu) = -{\dot\xi}^i(t) -\frac12 \bar{\epsilon}_+ \gamma_i \psi_{\slashed 0 -}(x^\nu)$\\[.1truecm]
\hline\hline
$m_\emptyset(x^\nu) = \Phi(x^\nu)\,,\ \ \omega_\emptyset{}^a(x^\nu) = -\partial^a\Phi(x^\nu) $&$ \psi_{\emptyset -}(x^\nu)=\Psi(x^\nu) $\\[.1truecm]
\hline
\end{tabular}
\caption{This table indicates the gauge fixing conditions and corresponding compensating transformations that lead to  3D Galilean  supergravity. We have also included the restrictions that follow from the fact that the spin-connection field $\omega_\mu{}^{ab}$ is dependent. At the bottom of the table we have summarized the expressions of the
non-zero remaining gauge fields in terms of the Newton potential $\Phi(x^\nu)$ and its supersymmetric partner
$\chi(x^\nu)$, which is related to $\Psi(x^\nu)$ via \eqref{defchi}.}
\end{center}
 \end{table}
 }

\section{Discussion}

In this work we constructed a supersymmetric extension of three-dimensional Newton-Cartan gravity
by gauging the $\mathcal{N}=2$ supersymmetric Bargmann algebra. An, at first sight,  un-usual feature we encountered is that
only half of the $\mathcal{N}=2$ supersymmetry is realized locally, the other half manifests itself
as a fermionic Stueckelberg symmetry. After fixing the Stueckelberg symmetry the second supersymmetry  is realized
only as a global supersymmetry.
A similar feature occurs in the bosonic case where the time reparametrizations occur as a Stueckelberg symmetry that after fixing leaves us with  constant time translations only.

We have discussed a full gauging, corresponding to `Newton-Cartan supergravity' and a medium gauging, obtained by partial gauge fixing, corresponding to `Galilean supergravity'. In the latter formulation, we have been able to realize the supersymmetry algebra on a multiplet containing the Newton potential, as well as its dual. The Newton potential and its dual correspond to the real and imaginary parts of a holomorphic function. This holomorphic structure is reminiscent of the three-dimensional relativistic case \cite{Deser:1983tn}, as well as of branes with two transverse directions such as cosmic strings and D7-branes \cite{Greene:1989ya,Gibbons:1995vg}. It would be interesting to see how these features can be generalized to higher dimensions.

The reason that we restricted ourselves to  three-dimensional Newton-Cartan supergravity   is that it is non-trivial
to find the additional fields, beyond the gauge fields associated to the supersymmetric Bargmann algebra,
that are needed to realize the supersymmetry algebra. This is different from the relativistic case where an off-shell counting of the field degrees of freedom restricts the possible choices. One way to make progress here is to better understand the representation theory of the super-Bargmann algebra thereby mimicking the relativistic case. Another useful approach could be to extend the work of \cite{Julia:1994bs} and approach the issue from a five-dimensional point of view. We note that the reduction of a 5D Poincar\'e multiplet to 4D gives an irreducible 4D $\mathcal{N}=2$ Poincar\'e multiplet {\sl plus} a $\mathcal{N}=2$ vector multiplet. It is not clear that
such a reducibility into two multiplets also occurs in the non-relativistic case. This might indicate that more fields,
namely those of the vector multiplet, are needed to close the supersymmetry algebra in the non-relativistic case\,\footnote{We thank Hermann Nicolai for a discussion on this point.}.

It is clear that more work needs to be done to come at a full grasp of the possible Newton-Cartan supergravities in arbitrary dimensions. We hope that this work, starting with the three-dimensional case, will help to better understand the higher-dimensional cases.

\section*{Acknowledgements} We thank the Simons Center for Geometry and Physics,
where an important part of this work was done during the 2012 Summer
Simons Workshop in Mathematics and Physics, for its hospitality,
financial support and stimulating ambiance. We thank Mees de Roo, Gary Gibbons, Joaquim Gomis, Marc Henneaux and Jan-Willem van Holten for discussions. E.B. wishes to thank
Hermann Nicolai for an interesting discussion on null reductions. The work of J.R. was supported by the START project Y 435-N16 of the Austrian Science Fund (FWF). The research of E.S. is supported in part by NSF grants PHY-0906222 and PHY-1214344.

\appendix

\section{Notation and Conventions}


Flat indices are denoted by capital Latin letters $A$, $B$,... In the Newton-Cartan formalism, they are split in time-like and space-like flat indices $\{0,a\}$. Curved indices are denoted by Greek letters $\mu$, $\nu$, ... and are split as $\mu=\{\emptyset,i\}$. Raising and lowering is still done using the usual Minkowski metric (with signature mostly plus). Raising or lowering a 0-index is thus done at the expense of a minus-sign. Turning curved into flat indices is done using the (inverse) vielbeins $\tau^\mu$ and $e^\mu{}_a$, as in the following example:
\begin{align}
\hat{F}_{0a} & =  \tau^\mu e^\nu{}_a \hat{F}_{\mu \nu} \,, \nonumber \\
\hat{F}_{ab} & =   e^\mu{}_a e^\nu{}_b \hat{F}_{\mu \nu} \,.
\end{align}

The relations \eqref{vielbrelSNC} can be used to turn flat into curved indices. They also imply the inverse vielbein variations
\begin{align}
\delta e^{\mu}{}_{a} & = - e^{\mu}{}_{b} e^{\nu}{}_{a} \delta e_{\nu}{}^{b} - \tau^{\mu}e^{\nu}{}_{a} \delta\tau_{\nu} \,, \\
\delta \tau^{\mu} & = -\tau^{\mu}\tau^{\nu}\delta \tau_{\nu} - e^{\mu}{}_{a} \tau^{\nu}\delta e_{\nu}{}^{a} \,. \label{variationsupperindexvielbeins}
\end{align}
To check the closure of the supersymmetry algebra on fermions, the following three-dimensional Fierz identity
\begin{align}
\psi\bar{\lambda}  = - \frac{1}{2}(\bar{\lambda}\psi) - \frac{1}{2}(\bar{\lambda} \gamma^0 \psi) \gamma_0 - \frac{1}{2}(\bar{\lambda} \gamma^a \psi) \gamma_a \label{Fierz1}
\end{align}
is needed.

\providecommand{\href}[2]{#2}\begingroup\raggedright
\endgroup

\end{document}